\documentclass[11pt]{article}
\usepackage{epsfig} 
\usepackage{amssymb}
\newcommand{\sect}[1]{ \section{#1} \setcounter{equation}{0} } 

\newcommand{\half}{\mbox{\small{$\frac{1}{2}$}}} 
\newcommand{\third}{\mbox{\small{$\frac{1}{3}$}}} 
 
\newcommand{\MSbar}{\overline{\mbox{MS}}} 

\newcommand{\Nf}{N_{\!f}}

\newcommand{\pitwo}{\mbox{\small{$\frac{\pi}{2}$}}}
\newcommand{\pisix}{\mbox{\small{$\frac{\pi}{6}$}}}

\newcommand{\pslash}{p \! \! \! /}
\newcommand{\qslash}{q \! \! \! /}

\begin{document}
\title{Symmetric point flavour singlet axial vector current renormalization at 
two loops}
\author{J.A. Gracey, \\ Theoretical Physics Division, \\ 
Department of Mathematical Sciences, \\ University of Liverpool, \\ P.O. Box 
147, \\ Liverpool, \\ L69 3BX, \\ United Kingdom.} 
\date{}

\maketitle 

\vspace{5cm} 
\noindent 
{\bf Abstract.} We calculate the two loop correction to the quark $2$-point
function with the non-zero momentum insertion of the flavour singlet axial
vector current at the fully symmetric subtraction point for massless quarks in
the modified minimal subtraction ($\MSbar$) scheme. The Larin method is used to
handle $\gamma^5$ within dimensional regularization at this loop order ensuring
that the effect of the chiral anomaly is properly included within the 
construction. 

\vspace{-16cm}
\hspace{13.2cm}
{\bf LTH 1228}

\newpage 

\sect{Introduction.}

One of the more curious experimental results over a generation ago was that of 
the EMC collaboration, \cite{1}. They measured the origin of the proton spin
and discovered that against expectations it was not due in a major part to the 
valence quarks. As the proton is a bound state of three quarks it was widely 
assumed that the combination of their quark spins would be the source of the 
overall spin-$\half$ of the proton. Instead the experiment observed that the 
gluons binding the quarks together give a sizeable contribution. This was 
surprising due to the fact that in some sense the gluons are sea partons. While
the original experiment was subsequently refined and improved to confirm the 
original observation, \cite{2,3,4,5}, a clear theoretical understanding was 
sought to explain the phenomenon. As such a venture requires the use of the 
strong sector of the Standard Model described by Quantum Chromodynamics (QCD), 
tools had to be developed and refined to tackle the problem. Moreover, to do so
one has to study an energy regime which is in the infrared and hence outside 
the region where perturbation theory is valid. Therefore the only viable 
approach was the application of lattice gauge theory which can access the 
non-perturbative structure of the proton through heavy use of supercomputers. 
Clearly such an exercise required new methods such as the inclusion of 
dynamical fermions and the field is better placed now to answer the theoretical
question of the source of the proton spin. This is not an isolated exercise for
lattice studies. As an aside it is worth mentioning that recently the breakdown
of the proton mass in terms of its constituent entities such as quark, gluon, 
weak sector and anomaly contributions has been accurately estimated on the 
lattice, \cite{6}. This entailed measuring the diagonal components of the 
energy-momentum tensor. Indeed the study given in \cite{6} has indicated that 
more accurate knowledge of the internal proton structure can be adduced 
theoretically in the near future. Parenthetically it is also worth noting the 
related problem of the pressure inside the proton. Experimentally this can be 
deduced accurately now as demonstrated in \cite{7}. In terms of theoretical 
studies and in particular those for the lattice such a pressure problem 
translates into requiring precise measurements of the off-diagonal components 
of the energy-momentum tensor. Such an exercise has been carried out recently 
in \cite{8,9}. However progress on studying the source of the proton spin on 
the lattice over the last few years can be seen in a non-exhaustive set of 
articles, \cite{10,11,12,13,14}, while a more detailed overview of this and the 
status of future directions of hadron physics computed on the lattice can be 
found in \cite{15,16,17,18}.

Concerning proton spin measurements on the lattice the quantum field theory 
formalism behind such potential calculations originate in the work of Ji, 
\cite{19}. There the relevant, in the sense of important, operators were 
identified and it was shown how their expectation values relate to the overall 
proton spin. Indeed central to the three properties of the proton mentioned 
already is the need to study matrix elements of various key operators for each 
observable. While the energy-momentum tensor provides the main operator in 
relation to mass and pressure, in the spin case it is the flavour singlet axial
vector current or $\bar{\psi} \gamma^5 \gamma^\mu \psi$ where $\psi$ and 
$\bar{\psi}$ are the respective quark and anti-quark of the same flavour. 
Indeed one of the motivations for studying the singlet axial vector operator is
that the difference between the singlet and non-singlet cases provides a way of 
quantifying the strange quark contribution to the proton spin, \cite{20,21,22}.
Over the years similar quark bilinear operators have been widely studied on the
lattice, \cite{20,23,24,25,26,27,28}. Aside from the associated matrix element 
which determines the non-perturbative structure, the operator renormalization 
has to be understood in the lattice regularized theory in order to implement 
the renormalization group running over momentum scales. In indicating the 
similar operators we mean the flavour non-singlet quark bilinear operators 
which are termed scalar, vector, tensor, axial vector and pseudoscalar 
depending on their Lorentz properties. Moreover, the matrix elements have been 
computed for a variety of configurations. These break into two classes known as
forward and non-forward where the former has the operator inserted at zero 
momentum in a quark $2$-point function. In the latter case it has a momentum 
flowing through it and the square of that momentum and those of the two 
external quarks can take different non-zero values. This non-forward set-up 
allows more freedom to probe detailed structure within nucleons. Most lattice 
studies of the flavour axial singlet current have been for the forward case, 
\cite{20,23,24,25,26,27,28}, and in the associated lattice renormalization 
scheme termed the modified regularization invariant (RI${}^\prime$) scheme 
introduced in \cite{29,30}.

One aspect of making lattice measurements in general is that of ensuring the
continuum limit is taken accurately. In recent years to assist this the process
has been adopted where the relevant matrix element is evolved to the 
ultraviolet region and matched to the continuum perturbative expansion of the 
same matrix element or Green's function. Clearly the more loop orders that are
available in the perturbative expansion means the matching will be more 
accurate and hence the lattice error estimates can be improved. For quark
bilinear operators the early work in this direction was provided in 
\cite{29,30}. In the context of lattice spin measurements there have been
studies, \cite{23,24,25,26,27,28,31} where the nucleon isovector scalar, axial 
vector and tensor charges were measured. For example \cite{31} built on an 
earlier parallel study in \cite{27} where the nucleon axial form factors were 
computed and the issues centered on the chiral anomaly taken into account 
through a nonperturbative treatment. However in the matching to the continuum 
in the work of \cite{31} only one loop information was available for the 
flavour singlet axial vector current. This is because at that loop order the 
matrix element for the flavour singlet and non-singlet axial vector currents 
are the same. The difference in these only occur at two loop order. The main 
reason for this is that the flavour singlet axial vector current is not 
conserved due to the chiral anomaly and its effect in the matrix element 
becomes present at two loops. In Feynman graph language there are graphs that 
are zero in the flavour non-singlet case but non-zero for the flavour singlet 
operator. By contrast in lattice language there are disconnected contributions 
to the proton correlation function when probed with the axial vector current in
the flavour singlet case. Therefore while the latter have been incorporated in 
the lattice simulations, \cite{27,31}, their omission in the matching to the 
continuum in \cite{31}, albeit due to taking only one loop data, means that the
effect of the chiral anomaly has not been taken into account. In other words 
the central values measured on the lattice for the hadronic matrix element will
not have accommodated the discrepancy in the continuum difference between the 
flavour non-singlet and singlet operators. This is due to the lack of a two 
loop computation of the relevant Green's function containing the flavour 
singlet axial vector current. 

Therefore it is the purpose of this article to close this particular gap. By 
doing so we will bring all the quark bilinear operator Green's functions to the
same loop level for the non-forward case. Specifically we will compute the two 
loop quark $2$-point function with the singlet axial vector current at non-zero
momentum insertion for the fully symmetric momentum configuration. Such a 
configuration is a non-exceptional one and hence should avoid infrared 
problems. The extension of the early lattice work on operator renormalization
at an exceptional point of \cite{29,30} was extended to the non-exceptional
case in \cite{32} at one loop and later to two loops in a variety of articles,
\cite{33,34,35,36,37}. More recently in the lattice study of \cite{38} the 
operator renormalization constants for all the flavour non-singlet quark 
bilinear operators were measured and it was demonstrated that for massless 
quarks those for the non-exceptional configuration were much more reliable in 
the infrared limit. Therefore in focussing on the fully symmetric point we are 
aiming at minimizing other potential sources of avoidable error for matching to
lattice data. While straightforward to state, the two loop computation we will 
undertake in dimensional regularization is also fraught with technical 
complications. One obvious one concerns the treatment of $\gamma^5$ but we will
use the Larin approach, \cite{39}, which is valid at least to the loop orders 
we are interested in. For practical multiloop computations the method adapted 
the early work of \cite{40,41,42} to incorporate $\gamma^5$ in dimensional 
regularization. Moreover, the chiral anomaly was correctly treated in that
approach beyond one loop. As part of our study we will extend the Larin 
construction in the sense that the non-forward matrix elements are computed. So
we will show how the {\em same} finite renormalization constant emerges as that
of \cite{39} to ensure chiral symmetry is correctly present after 
renormalization and the subsequent lifting of the dimensional regularization. 
This is a non-trivial task and in some sense the study again substantiates the
foresight and elegance of the work of \cite{40,41,42}. In saying this we will 
also check the same finite renormalization constant arises for the pseudoscalar
current in the non-forward configuration thereby verifying that a consistent 
picture emerges. En route we will discuss a minor modification of the Larin 
approach for flavour non-singlet operators. In \cite{39} the criterion to 
define the finite renormalization constant needed to ensure four dimensional 
properties correctly emerge for the two spin-$1$ (non-singlet) quark bilinear 
operators was to implement equality of the two relevant Green's functions. We
show that this can also be achieved from ensuring the currents are conserved in
four dimensions after renormalization.

The article is organized as follows. In the following section we introduce the
formalism used to carry out the two loop computation as well as discuss the
treatment of $\gamma^5$ in dimensional regularization using the Larin method,
\cite{39}. Included in this section is the algorithm we implement to evaluate 
the singlet axial vector current Green's function at the symmetric point. 
Section $3$ is devoted to the discussion of our results where we quantify the 
difference between the flavour non-singlet and singlet axial vector operator
Green's functions prior to providing concluding remarks in Section $4$. 

\sect{Background.}

We devote this section to describing the details of the computation and en
route review previous renormalizations of the operators in question at the
symmetric point. In order to appreciate the subtleties between the flavour 
non-singlet and singlet quark bilinear operators we consider we define the
non-singlet operators as
\begin{equation}
{\cal S}^{ns} ~ \equiv ~ \bar{\psi}^i \psi^j ~~~,~~~
{\cal V}^{ns} ~ \equiv ~ \bar{\psi}^i \gamma^\mu \psi^j ~~~,~~~
{\cal T}^{ns} ~ \equiv ~ \bar{\psi}^i \sigma^{\mu\nu} \psi^j ~~~,~~~
{\cal A}^{ns} ~ \equiv ~ \bar{\psi}^i \gamma^5 \gamma^\mu \psi^j ~~~,~~~
{\cal P}^{ns} ~ \equiv ~ \bar{\psi}^i \gamma^5 \psi^j 
\label{oplabel}
\end{equation}
and 
\begin{equation}
{\cal S}^s ~ \equiv ~ \bar{\psi}^i \psi^i ~~~,~~~
{\cal V}^s ~ \equiv ~ \bar{\psi}^i \gamma^\mu \psi^i ~~~,~~~
{\cal T}^s ~ \equiv ~ \bar{\psi}^i \sigma^{\mu\nu} \psi^i ~~~,~~~
{\cal A}^s ~ \equiv ~ \bar{\psi}^i \gamma^5 \gamma^\mu \psi^i ~~~,~~~
{\cal P}^s ~ \equiv ~ \bar{\psi}^i \gamma^5 \psi^i 
\label{oplabels}
\end{equation}
for the singlet case where $i$ and $j$ are flavour indices and there is no sum
over $i$ in the latter set. Given that our main interest is in the perturbative
structure of a specific Green's function for a particular external momentum
configuration we define the general case, which includes the above operators as
well, by
\begin{equation}
\Gamma^{{\cal O}^{\cal I}} ~=~ \left. \left\langle \frac{}{} 
\psi(p) {\cal O}^{\cal I}(-p-q) \bar{\psi}(q) \right\rangle 
\right|_{p^2=q^2=(p+q)^2=-\mu^2} 
\label{opgf}
\end{equation}
where the operator ${\cal O}$ corresponds to any one of (\ref{oplabel}) and 
(\ref{oplabels}). We use a similar notation to \cite{35,36,37} and follow the
general approach provided there. The two independent external momenta, $p$ and 
$q$, satisfy the condition for the symmetric subtraction point which is, 
\cite{43,44},
\begin{equation}
p^2 ~=~ q^2 ~=~ ( p + q )^2 ~=~ -~ \mu^2 ~~~,~~~ pq ~=~ \frac{1}{2} \mu^2
\label{symmpt}
\end{equation}
where $\mu$ is a mass scale. This setup is a nonexceptional configuration and
therefore there are no infrared issues. The Green's function (\ref{opgf}) is 
illustrated in Figure \ref{figop} where an operator ${\cal O}^{\cal I}$ of 
(\ref{oplabel}) or (\ref{oplabels}) is indicated by the crossed circle. 

{\begin{figure}[ht]
\begin{center}
\includegraphics[width=7.0cm,height=7.0cm]{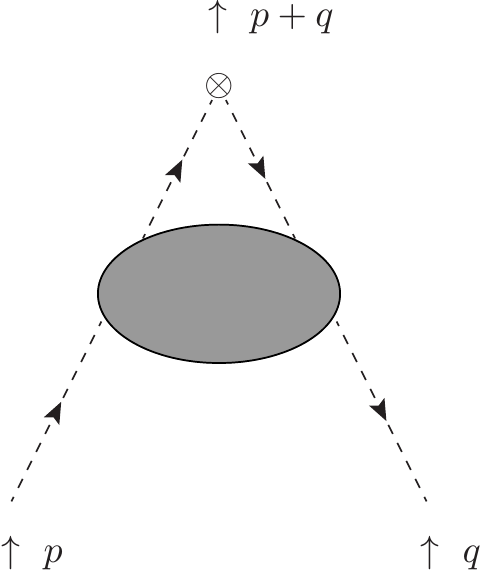} \\
\end{center}
\caption{Momentum configuration for $\left\langle \psi(p) {\cal O}^i (-p-q)
\bar{\psi}(q) \right\rangle$.}
\label{figop}
\end{figure}}

Having defined the Green's function the first step in its perturbative
evaluation is the generation of the two loop Feynman graphs. To do this we
have used the {\sc Qgraf} package, \cite{45}, which produces $1$ one loop and
$13$ two loop graphs. At the former order the expressions for the respective
flavour non-singlet and singlet Green's functions are the same irrespective of
the subtraction point. The first place where any discrepancy will appear as a 
result of flavour symmetry in the chiral limit is at two loops and is due to
the two graphs of Figure \ref{fig2l}. This is because these graphs are zero
for the flavour non-singlet case as the insertion of such an operator into the
closed fermion loop gives a trace over a traceless flavour group generator. So 
such graphs will not contribute to $\Gamma^{{\cal I}^{ns}}$. By contrast for
the flavour singlet case the graphs of Figure \ref{fig2l} will not be zero by
this particular flavour trace argument. However for the operators ${\cal S}^s$,
${\cal T}^s$ and ${\cal P}^s$ the graphs of Figure \ref{fig2l} are zero since 
there are an odd number of $\gamma$-matrices in the closed loop in the chiral
limit. These graphs would correspond to the disconnected graphs in the proton
correlation function on the lattice. For ${\cal V}^s$ and ${\cal A}^s$ there 
will be an even number of $\gamma$-matrices in the loop in the massless case 
considered here. So these are the two possible instances of the flavour 
symmetry producing different two loop Green's functions (\ref{opgf}). As we 
will be using dimensional regularization in our calculations one concern that 
could be raised is whether this argument applies for operators containing 
$\gamma^5$. In the case of ${\cal A}^{ns}$ and ${\cal P}^{ns}$ we are permitted
to use the naive anticommuting $\gamma^5$, \cite{40,41,42}, since it always 
appears in Feynman integrals in an open string of $\gamma$-matrices. It is its 
presence, or an odd number of them, within a closed fermion loop that requires 
special treatment in dimensional regularization, \cite{40,41,42}. We will 
discuss this in detail later aside from noting that when $\gamma^5$, or an odd 
number of them, is present in a closed loop with an odd number of ordinary 
$\gamma$-matrices then the spinor trace is still zero as in four dimensions. 
Having taken flavour and Lorentz symmetries into account there is one final 
constraint to be considered however which is that corresponding to the colour 
vector space. All operators of (\ref{oplabel}) and (\ref{oplabels}) are colour 
singlets. In other words they do not include a colour group generator. So the 
colour trace is not the same as the parallel graphs contributing to the 
quark-gluon vertex function. In that instance if the operator insertion was 
replaced by a gluon then the sum of the fermion one loop subgraphs would be 
proportional to the structure constants $f^{abc}$. In the case of ${\cal V}^s$ 
the colour trace with one fewer group generator produces $\delta^{ab}$ which 
together with the relative minus sign arising from the respective 
$\gamma$-matrix traces means that the two graphs of Figure \ref{fig2l} sum to 
zero for this operator. For ${\cal A}^s$ by contrast while the colour argument 
applies equally, the presence of $\gamma^5$ in the spinor trace prevents the 
same procedure giving zero since the traces sum and are not equal and opposite.
Therefore the only singlet operator that needs to be considered at any 
subtraction point, and not just the symmetric one, is ${\cal A}^s$ as it will 
produce contributions additional to its non-singlet partner.

{\begin{figure}[ht]
\begin{center}
\includegraphics[width=9.5cm,height=4.5cm]{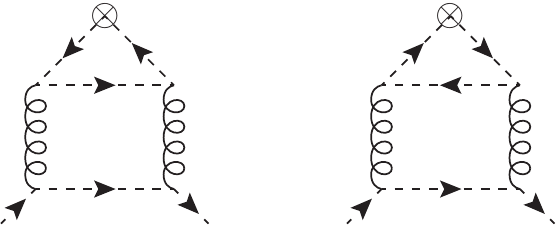} \\
\end{center}
\caption{Extra graphs for flavour singlet operators.}
\label{fig2l}
\end{figure}}

Having isolated the only case we have to consider for singlet operators then in
order to evaluate $\Gamma^{{\cal A}^s}$ we note that we have taken a different
path to the partner computation of $\Gamma^{{\cal A}^{ns}}$ of 
\cite{35,36,37}. To appreciate the contrast it is first best to summarize the 
earlier approach. In \cite{35,36,37} a projection method was used whereby the 
Green's function was decomposed into a basis of tensors consistent with the 
symmetries of the operator. For instance, for ${\cal V}$ this was   
\begin{eqnarray}
{\cal P}^{\cal V}_{(1) \mu }(p,q) &=& \gamma_\mu ~~,~~
{\cal P}^{\cal V}_{(2) \mu }(p,q) ~=~ \frac{p^\mu \pslash}{\mu^2} ~~,~~
{\cal P}^{\cal V}_{(3) \mu }(p,q) ~=~ \frac{p_\mu \qslash}{\mu^2} ~, 
\nonumber \\
{\cal P}^{\cal V}_{(4) \mu }(p,q) &=& \frac{q_\mu \pslash}{\mu^2} ~~,~~
{\cal P}^{\cal V}_{(5) \mu }(p,q) ~=~ \frac{q_\mu \qslash}{\mu^2} ~~,~~
{\cal P}^{\cal V}_{(6) \mu }(p,q) ~=~ \Gamma_{(3) \, \mu p q} 
\frac{1}{\mu^2} 
\label{basisvec}
\end{eqnarray}
and this Lorentz basis for each bilinear operator Green's function is the same 
whether the operator is flavour non-singlet or singlet. Here we have introduced
the generalized $\gamma$-matrices of \cite{42,46,47,48,49,50} which are denoted 
by $\Gamma_{(n)}^{\mu_1\ldots\mu_n}$ and defined by
\begin{equation}
\Gamma_{(n)}^{\mu_1 \ldots \mu_n} ~=~ \gamma^{[\mu_1} \ldots \gamma^{\mu_n]}
\label{gamn}
\end{equation}
for integers $n$ with $0$~$\leq$~$n$~$<$~$\infty$. Each matrix is fully 
antisymmetric in the Lorentz indices for $n$~$\geq$~$2$ and the full set span 
the infinite dimensional spinor space when dimensional regularization is 
implemented. We note that $\Gamma_{(5)}^{\mu_1\mu_2\mu_3\mu_4\mu_5}$ is not
related to $\gamma^5$ and in strictly four dimensions 
$\Gamma_{(n)}^{\mu_1\ldots\mu_n}$~$=$~$0$ for $n$~$\geq$~$5$. One advantage of 
these matrices is that they provide a natural partition since
\begin{equation}
\mbox{tr} \left( \Gamma_{(m)}^{\mu_1 \ldots \mu_m}
\Gamma_{(n)}^{\nu_1 \ldots \nu_n} \right) ~ \propto ~ \delta_{mn}
I^{\mu_1 \ldots \mu_m \nu_1 \ldots \nu_n}
\end{equation}
with $I^{\mu_1 \ldots \mu_m \nu_1 \ldots \nu_n}$ denoting the unit matrix in
$\Gamma$-space.
Here we use the convention that when a momentum is contracted with a Lorentz
index in $\Gamma_{(n)}^{\mu_1\ldots\mu_n}$ then the momentum itself appears in 
place of the index. 

Once the basis for the Green's function has been chosen then the coefficients
of each tensor is deduced by multiplying it by a $d$-dimensional linear
combination of the tensors to produce a sum of scalar Feynman integrals. These
are then evaluated by applying the Laporta algorithm, \cite{51}, which 
systematically integrates by parts all the contributing graphs. This produces a
linear combination of a small set of master integrals with $d$-dependent 
rational polynomial coefficients. For the symmetric point configuration 
explicit expressions for the one and two loop master integrals have been 
available for many years, \cite{52,53,54,55}. Though in more recent years they 
have been understood in the language of cyclotomic polynomials, \cite{56}. To 
be explicit to two loops the Green's functions with the kinematic configuration
of (\ref{symmpt}) involve different linear combinations of numbers from the set
\begin{equation}
\left\{ \mathbb{Q}, \pi^2, \zeta_3, \zeta_4,
\psi^\prime( \third ),
\psi^{\prime\prime\prime}( \third ),
s_2( \pitwo ),
s_2( \pisix ), 
s_3( \pitwo ),
s_3( \pisix ),
\frac{\ln^2(3) \pi}{\sqrt{3}},
\frac{\ln(3) \pi}{\sqrt{3}},
\frac{\pi^3}{\sqrt{3}}
\right\} ~.
\label{numbas}
\end{equation}
Here $\zeta_n$ is the Riemann zeta-function, $s_2(z)$ and $s_3(z)$ are defined 
in terms of the polylogarithm function $\mbox{Li}_n(z)$ 
\begin{equation}
s_n(z) ~=~ \frac{1}{\sqrt{3}} \Im \left[ \mbox{Li}_n \left( 
\frac{e^{iz}}{\sqrt{3}} \right) \right]
\end{equation}
and $\psi(z)$ is the derivative of the logarithm of the Euler 
$\Gamma$-function. A final important aspect of the computation was the 
extensive use of symbolic manipulation which was facilitated through the use of
the language {\sc Form} and its threaded version {\sc Tform}, \cite{57,58}. 
Using the {\sc Reduze} package, \cite{59,60}, written in C$++$ which implements
the Laporta reduction we inserted the relations generated by the package for 
the required Feynman integrals via a {\sc Form} module so that all our 
computations were carried out automatically. In addition we followed the 
prescription of \cite{61} to implement the operator renormalization within the
same automatic framework. 

As we will be considering Green's functions involving operators with $\gamma^5$
present we need to be careful in its treatment within dimensional
regularization. Moreover the strategy we will choose to follow has to be robust
in the sense that it should reproduce the known two loop symmetric point
results of earlier work, \cite{32,33,34,35}. For the non-singlet example of 
${\cal A}^{ns}$ and ${\cal P}^{ns}$ there are several ways one can proceed. 
First though we need to discuss the details of the Larin approach, \cite{39},
as general features of that computation will be used. It was inspired by and 
developed from the earlier one loop resolution of the treatment of $\gamma^5$ 
of \cite{40,41,42,62}. In reviewing \cite{39} it is important to note that 
those calculations were carried out at an exceptional momentum configuration 
where the operator is inserted at zero momentum. In this case and also for the 
symmetric point, which is non-exceptional, for open strings of 
$\gamma$-matrices which include $\gamma^5$ matrices their naive 
anti-commutation with $d$-dimensional $\gamma$-matrices is valid. Related to 
this in \cite{63} an extended set of (non-singlet) quark bilinear operators 
were considered in $d$-dimensions and renormalized in the dimensionally 
regularized theory. These were
\begin{equation}
{\cal O}_{(n)}^{\mu_1\ldots\mu_n\,{ns}} ~=~ 
\bar{\psi}^i \Gamma_{(n)}^{\mu_1\ldots\mu_n} \psi^j 
\label{genop}
\end{equation}
and in four dimensions they reduce to ${\cal S}^{ns}$, ${\cal V}^{ns}$ and
${\cal T}^{ns}$ respectively for $n$~$=$~$0$, $1$ and $2$. For $n$~$=$~$3$ and
$4$ we have 
\begin{equation}
\left. {\cal O}_{(3)}^{\mu\nu\sigma\,{ns}} \right|_{d=4} ~=~ 
\epsilon^{\mu\nu\sigma\rho} \bar{\psi}^i \gamma^5 \gamma_\rho \psi^j ~~~,~~~ 
\left. {\cal O}_{(4)}^{\mu\nu\sigma\rho\,{ns}} \right|_{d=4} ~=~ 
\epsilon^{\mu\nu\sigma\rho} \bar{\psi}^i \gamma^5 \psi^j
\label{genoppseu}
\end{equation}
where $\epsilon^{\mu\nu\sigma\rho}$ is the totally antisymmetric strictly four
dimensional pseudo-tensor. Like $\gamma^5$ it has no existence outside strictly
four dimensions. For $n$~$\geq$~$5$ the operators of (\ref{genop}) are 
evanescent in the sense that they are not present in strictly four dimensions 
due to their being more free Lorentz indices than the spacetime dimension which
cannot be possible for a fully antisymmetric object. 

Before concentrating on the technical details of the application of \cite{39}
to the operators we are interested in here, it is worth detailing the early 
treatment of $\gamma^5$. The problem of accommodating an object, such as 
$\gamma^5$ that only exists in strictly {\em four} dimensions, within 
dimensional regularization was recognised in the seminal work of \cite{40}. In 
particular a formalism was developed in \cite{40} that consistently took into 
account the analytic continuation of the spacetime dimension to a complex 
variable $d$ together with the algebraic properties of $\gamma^5$ that are only
applicable in the underlying four dimensional physical subspace. The essence of
the approach of \cite{40} was to partition the $d$-dimensional spacetime into a 
physical four dimensional spacetime and a $(d-4)$-dimensional unphysical 
subspace. Purely four dimensional objects can only be elements of the former. 
Clear examples are $\gamma^5$ and $\epsilon^{\mu\nu\sigma\rho}$, which are 
present in (\ref{genoppseu}), whose indices can only take values in the 
physical subspace. One test of this construction was the successful 
verification, \cite{40}, of the one loop axial vector anomaly of 
\cite{64,65,66}. The full mathematical foundation for the treatment of 
$\gamma^5$ and $\epsilon^{\mu\nu\sigma\rho}$ in this partitioned spacetime was 
subsequently established in depth in \cite{62}. While such a method can readily
be applied at one loop level, effecting it in higher loop computations can only
proceed in a reasonable amount of time and in practice through the use of 
automatic Feynman diagram computation. Enacting such an approach turns out to 
be difficult but an effective algorithm that equates to this procedure and, 
moreover can be encoded, was developed in \cite{39} based on the ground work of
\cite{41,42,62}. In \cite{67,68} the definition of $\gamma^5$ in operators and 
currents adapted that introduced in \cite{40,41,42,62} and its relation to the 
renormalization procedure introduced in \cite{41} was explored. In particular 
the additional finite renormalization that is necessary to ensure purely four 
dimensional symmetries are respected in the resulting finite theory after the 
regularization has been lifted, was determined to high loop order. 

In light of this overview we now summarize its practical application to the 
multiplicatively renormalizable operators of interest, (\ref{genop}), at the
symmetric point. Given the relation (\ref{genoppseu}) between the generalized 
operators and their four dimensional counterparts, there is a connection with 
not only the renormalization of all the operators of (\ref{genop}) but 
importantly the respective Green's functions themselves should be in full 
agreement in strictly {\em four} dimensions after renormalization, \cite{41}. 
In other words the relation of (\ref{genoppseu}) ought to be valid in the 
renormalized theory for any scheme. This algorithm of \cite{41} was 
substantiated in \cite{39} in the modified minimal subtraction ($\MSbar$) 
scheme through a two stage process. The first part was to renormalize the 
operators ${\cal O}_{(3)}^{\mu\nu\sigma\,{ns}}$ and
${\cal O}_{(4)}^{\mu\nu\sigma\rho\,{ns}}$, for example, separately in the usual
way in the $\MSbar$ scheme to produce what is termed the naive renormalization 
constant for the $d$-dimensional operator. Aside from the one loop scheme 
independent term of the operator anomalous dimension the renormalization 
constants for the respective pairs ${\cal O}_{(0)}^{{ns}}$ and 
${\cal O}_{(4)}^{\mu\nu\sigma\rho\,{ns}}$, and ${\cal O}_{(1)}^{\mu\,{ns}}$ and
${\cal O}_{(3)}^{\mu\nu\sigma\,{ns}}$ disagreed. We note that while 
$\Gamma_{(1)}^\mu$ and $\gamma^\mu$ are equivalent we will retain the former 
notation when we discuss the renormalization of the set of operators of the 
form (\ref{genop}). This disagreement in the naive renormalization constants is
clearly not consistent with expectations from the naive anticommutation of 
$\gamma^5$ in {\em four} dimensions, \cite{40,41,62}. However the second stage
of the process developed in \cite{41} is to recognise that it is not possible 
to retain the symmetry properties of a purely four dimensional entity, 
$\gamma^5$, in the dimensionally regularized theory. So to circumvent the 
absence of the four dimensional properties in the dimensionally regularized 
theory, one has to augment the naive renormalization of 
${\cal O}_{(3)}^{\mu\nu\sigma\,{ns}}$ and
${\cal O}_{(4)}^{\mu\nu\sigma\rho\,{ns}}$ by including a finite renormalization 
constant pertinent to each operator, \cite{41}. The condition used to define 
this in \cite{39,62} for the flavour non-singlet case was to impose the 
constraint that after the naive renormalization of the Green's function with 
the $n$-index operator insertion the result is equivalent to that when $n$ is 
replaced by $(4-n)$, \cite{39,62}. Here $4$ is chosen as it is the critical 
dimension of QCD. There is a practical caveat to this in that the Lorentz 
tensor basis has to be written in terms of purely four dimensional objects 
{\em first} before the finite renormalization can be made. Also the finite 
renormalization constant in the $\MSbar$ scheme derives from the basis tensor 
corresponding to the tree term of the Green's function.  

For the fully symmetric point case we consider here we have adapted this 
approach but first checked that the previous non-singlet results are reproduced
for $n$~$=$~$3$ and $4$. However for the Green's functions of both operators 
the basis of tensors is larger than at the exceptional point. As a first step 
for $n$~$=$~$3$ and $4$ we have taken a slightly different tack to 
\cite{35,36,37} and avoided using a projection method on the Green's function. 
Instead to evaluate the contributing Feynman graphs we first removed all the 
$\gamma$-algebra from the Feynman integrals and evaluated the underlying tensor
integrals. One reason for this is that it bypasses the complication of 
constructing a decomposition into a Lorentz basis with $3$ and $4$ free indices
in the case of ${\cal A}^{ns}$ and ${\cal P}^{ns}$ respectively. In the latter 
case for instance that basis would include 
$\Gamma_{(6)}^{\mu\nu\sigma\rho p q}$ as one example. While this is evanescent 
it would have to be included to ensure the tensor basis was complete. So in 
avoiding a direct projection and consequently reproducing the same results as 
previous computations, \cite{32,33,34,35}, will ensure that we have established
a valid algorithm for handling $\gamma^5$ in the non-singlet case. This 
therefore will eventually be our strategy for the vector and axial vector 
operators in the {\em singlet} case when the complications due to the graphs of 
Figure \ref{fig2l} have to be taken into account. While the correct tensor 
basis will emerge from this integral projection the four dimensional basis of
(\ref{basisvec}) will not be the relevant one for the {\em naive} 
renormalization of ${\cal A}^{ns}$. Using relations such as
\begin{equation}
\Gamma_{(3)}^{\mu\nu\sigma} ~=~ \epsilon^{\mu\nu\sigma\rho} \gamma^5 
\gamma_\rho ~~~,~~~
\epsilon^{\mu\nu p q} \Gamma_{(1)\nu} ~=~ \gamma^5 \Gamma_{(3)}^{\mu p q}
\end{equation}
in four dimensions, for instance, the analogous basis will then be 
\begin{eqnarray}
{\cal P}^{\cal A}_{(1) \mu }(p,q) &=& \gamma^5 \gamma_\mu ~~,~~
{\cal P}^{\cal A}_{(2) \mu }(p,q) ~=~ \gamma^5 \pslash p^\mu 
\frac{1}{\mu^2} ~~,~~
{\cal P}^{\cal A}_{(3) \mu }(p,q) ~=~ \gamma^5 \qslash p_\mu 
\frac{1}{\mu^2} ~, 
\nonumber \\
{\cal P}^{\cal A}_{(4) \mu }(p,q) &=& \gamma^5 \pslash q_\mu 
\frac{1}{\mu^2} ~~,~~
{\cal P}^{\cal A}_{(5) \mu }(p,q) ~=~ \gamma^5 \qslash q_\mu 
\frac{1}{\mu^2} ~~,~~
{\cal P}^{\cal A}_{(6) \mu }(p,q) ~=~ \gamma^5 \Gamma_{(3) \, p q \mu}
\frac{1}{\mu^2} ~.~~~~
\label{basisvec5}
\end{eqnarray}
Next we recall that to properly renormalize the operator ${\cal V}^{ns}$ 
requires some care. As the non-singlet vector current is conserved in the 
chiral limit 
\begin{equation}
\partial_\mu \left( \bar{\psi}^i \gamma^\mu \psi^j \right) ~=~ 0
\label{vconns}
\end{equation}
its renormalization constant is unity to all orders in perturbation theory and
in {\em all} renormalization schemes. In the $\MSbar$ scheme context this means
that the Green's function for ${\cal V}^{ns}$ is completely finite. However in 
a kinematic renormalization scheme where renormalization constants can have
non-zero finite parts the renormalization constant for ${\cal V}^{ns}$ could
mistakenly be chosen by demanding that the coefficient of  
${\cal P}^{\cal V}_{(1) \mu }(p,q)$ is unity for instance. This would clearly 
contradict the general result that physical operators do not get renormalized. 
In practical terms the conservation of the currents is encoded in the quantum 
theory by a Ward-Takahashi identity. In the case of ${\cal V}^{ns}$ this 
corresponds to Green's function of 
$\partial_\mu \left( \bar{\psi}^i \gamma^\mu \psi^j \right)$ being
related to the quark $2$-point function. Therefore we have checked that for our
integral projection construction that this is indeed the case in the $\MSbar$
scheme for the fully symmetric and hence non-exceptional configuration. We note
that we found full agreement with the one and two loop results of 
\cite{32,33,34,35}. This was also checked in \cite{63} for the direct 
projection on the operator Green's function. 

Since the non-singlet axial vector current is also conserved 
\begin{equation}
\partial_\mu \left( \bar{\psi}^i \gamma^5 \gamma^\mu \psi^j \right) ~=~ 0
\label{avconns}
\end{equation}
similar general reasoning also applies. Calculating the Green's function of 
$\partial_\mu \left( \bar{\psi}^i \gamma^5 \gamma^\mu \psi^j \right)$,
however, it does not agree with $\gamma^5$ times the quark $2$-point function. 
This is consistent with Larin's observation that treating the naive 
renormalization of the generalized operator ${\cal O}_{(3)}^{\mu\nu\sigma\,ns}$
as being equivalent to ${\cal A}^{ns}$ is incorrect. Instead an additional 
finite renormalization constant is required to ensure the consistency with 
symmetry properties of the strictly four dimensional theory. Ensuring the
consistency with the quark $2$-point function in this case we find that to two 
loops at the symmetric point the finite renormalization is
\begin{equation}
Z^{\mbox{\footnotesize{fin}}}_{{\cal A}^{ns}} ~=~ 1 ~-~ 4 C_F a ~+~ \left[
22 C_F^2 - \frac{107}{9} C_F C_A + \frac{4}{9} C_F T_F \Nf \right] a^2 ~+~
O(a^3)
\label{zfinans}
\end{equation} 
in full agreement with \cite{39} where $a$~$=$~$g^2/(16\pi^2)$ and $g$ is the
gauge coupling constant. It should be stressed that we have verified that the 
finite renormalization is independent of the subtraction point which is a 
non-trivial observation. 

{\begin{figure}[ht]
\begin{center}
\includegraphics[width=7.7cm,height=7.7cm]{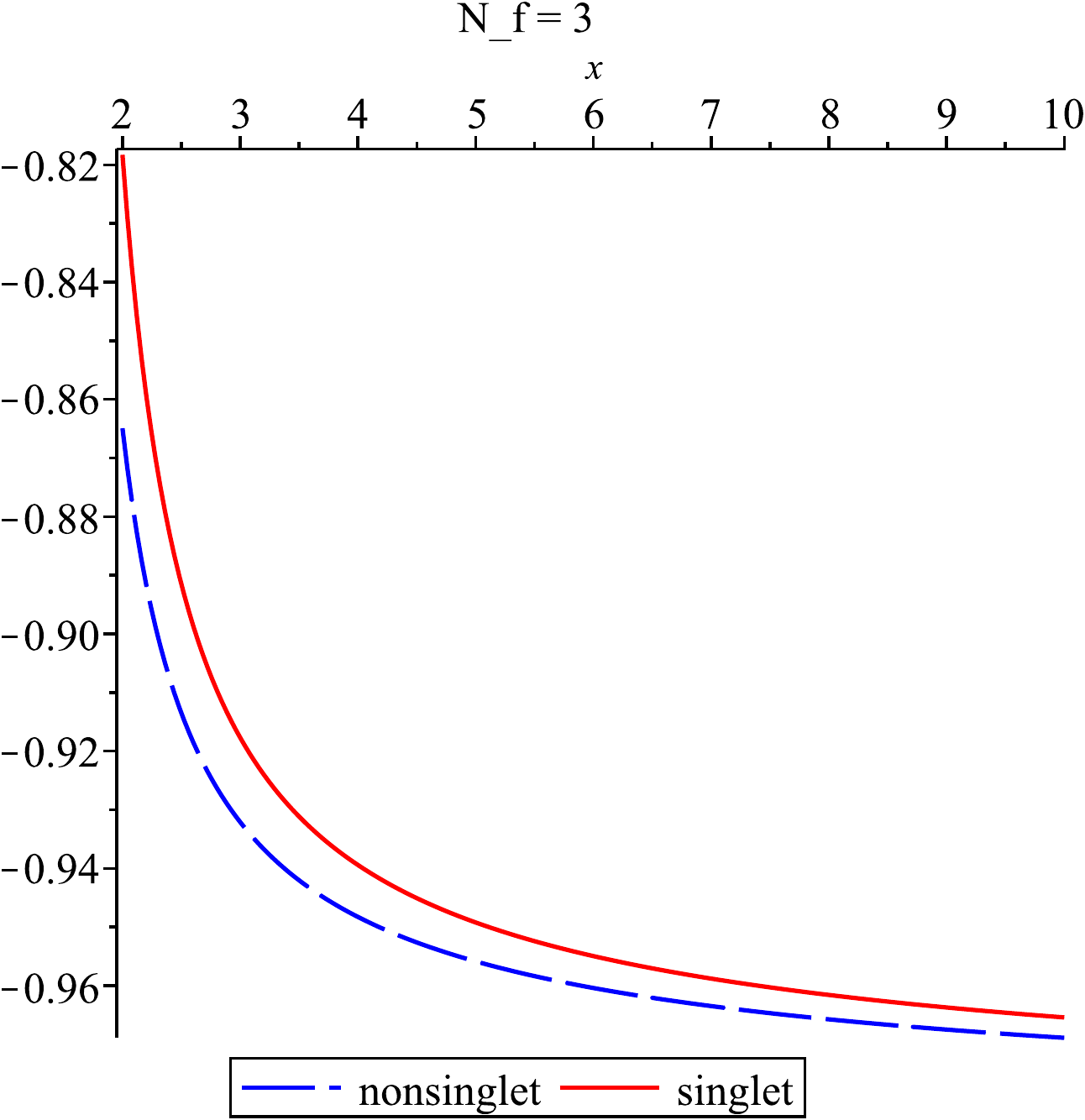}
\quad
\includegraphics[width=7.7cm,height=7.7cm]{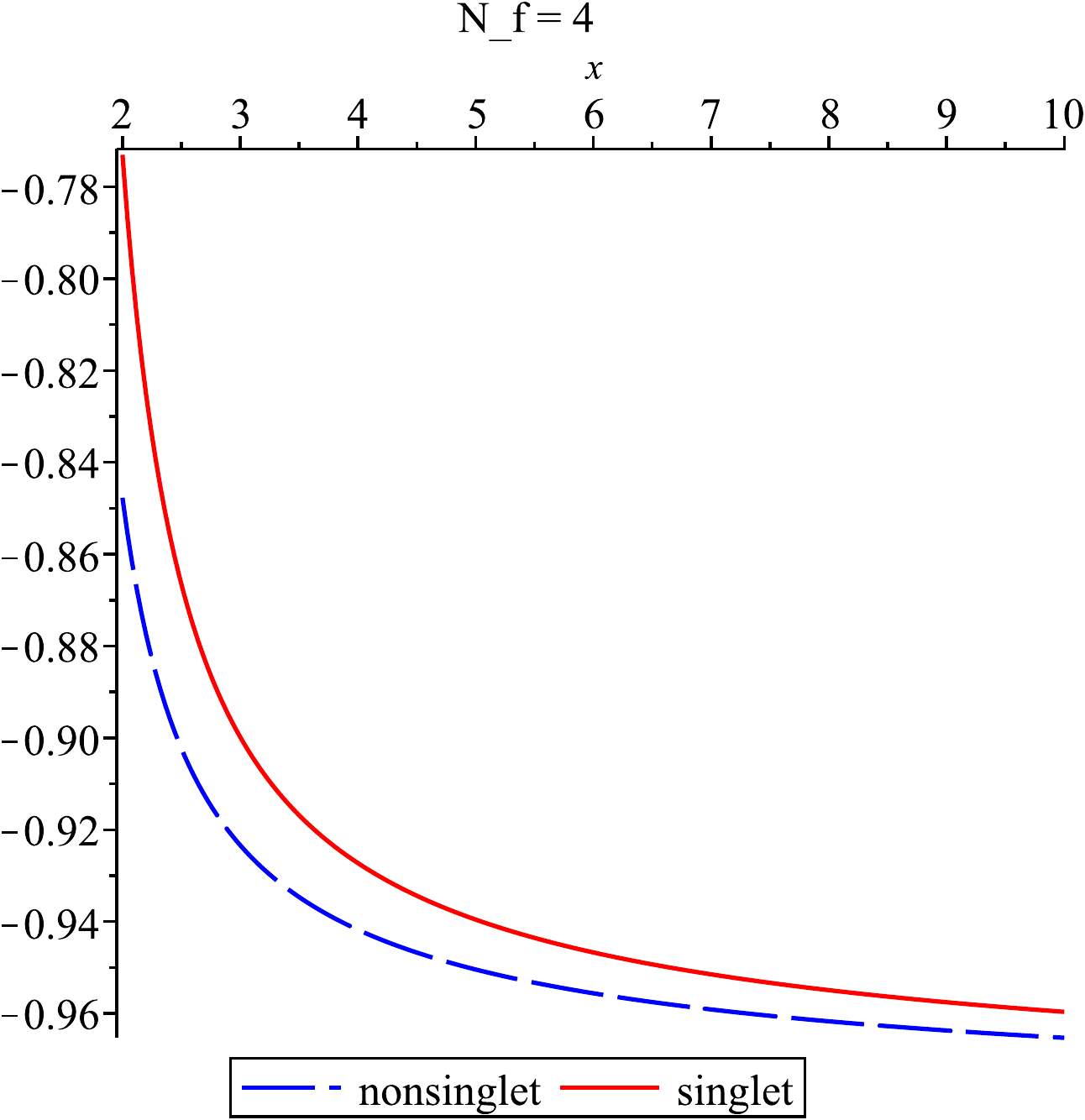}

\vspace{0.8cm}
\includegraphics[width=7.7cm,height=7.7cm]{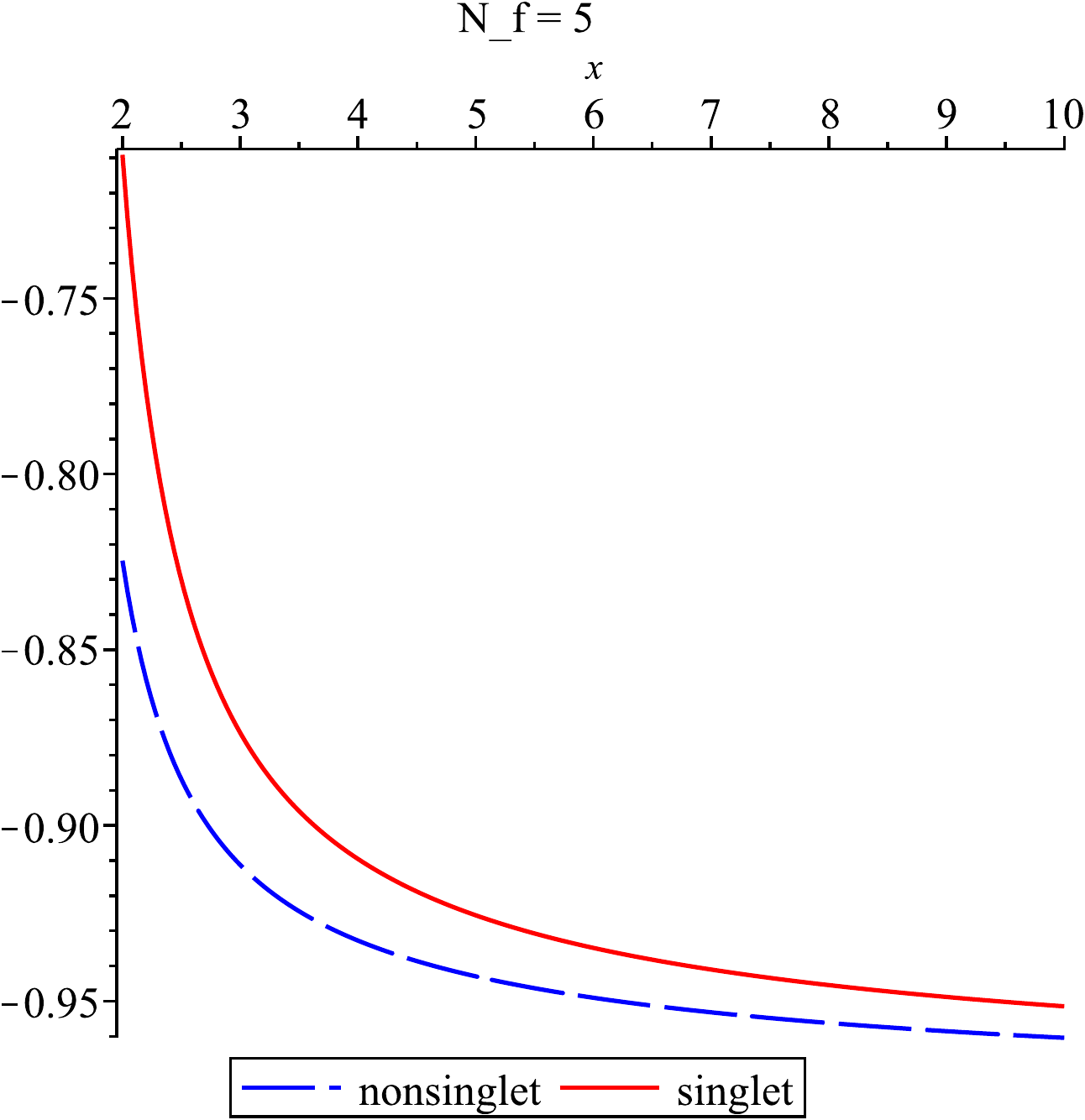}
\quad
\includegraphics[width=7.7cm,height=7.7cm]{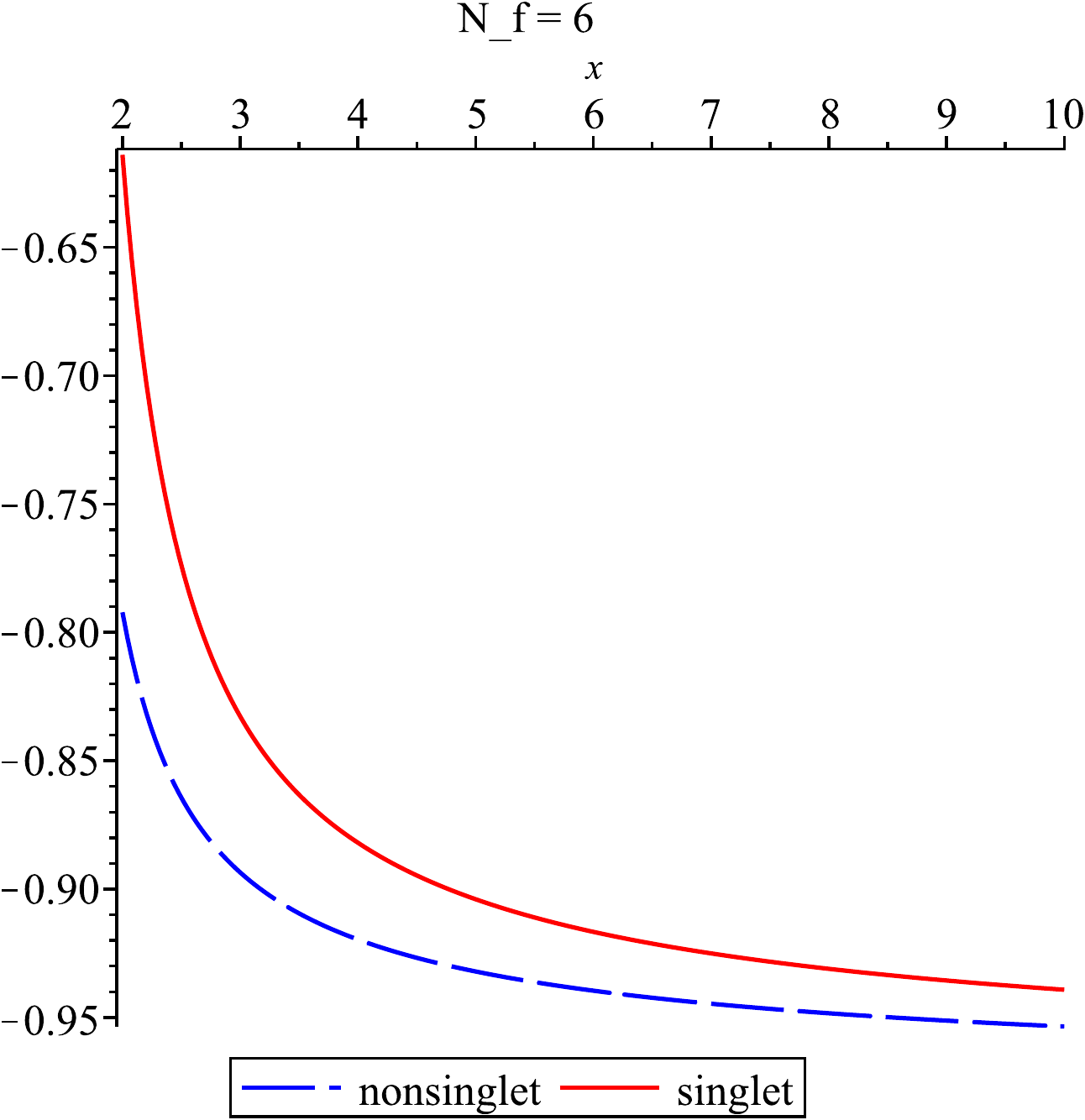}

\end{center}
\caption{Comparison of two loop coefficient of $\gamma^5\gamma^\mu$ in
$\Gamma^{{\cal A}^{ns}}$ and $\Gamma^{{\cal A}^s}$ for $\Nf$~$=$~$3$ to $6$.}
\label{figcomp}
\end{figure}}

As can be seen in earlier work \cite{32,34,35,36,37} the two loop explicit 
expression for (\ref{opgf}) for each operator involves linear combinations of 
(\ref{numbas}). These in principle could have been present in the two loop 
finite renormalization constant with our choice of (\ref{symmpt}). In the 
momentum configuration used in \cite{39} only rational numbers or $\zeta_3$ 
were present in the finite parts of the Green's function at three loops. So it 
is reassuring that the polylogarithms, for instance, of (\ref{numbas}) are 
absent. We note that although the three loop finite renormalization is also 
known in the $\MSbar$ scheme, \cite{39}, in order to verify the next term of 
(\ref{zfinans}) at the symmetric point would require a new three loop 
computation. At present the three loop master integrals are not available in 
order to be able to perform this calculation. This comparison of the 
$n$~$=$~$1$ and $3$ Green's function is in keeping with the spirit of 
\cite{39,62,63}. However in the singlet case the presence of the extra graphs 
of Figure \ref{fig2l} will not make this a viable way to proceed as was 
indicated in \cite{39}. In the non-singlet case the corresponding 
Ward-Takahashi identity for the axial vector operator provides a more field 
theoretic alternative method to determine the finite renormalization. In other 
words the Green's function of 
$\partial_\mu \left( \bar{\psi}^i \gamma^5 \gamma^\mu \psi^j \right)$ has to be
equivalent to the quark $2$-point function multiplied by $\gamma^5$. To ensure 
this we have checked that the same gauge parameter independent finite 
renormalization constant, (\ref{zfinans}), is required. It should also be noted
that we have checked that once this finite renormalization is determined from 
ensuring that the current conservation is preserved then the expressions for 
both Green's functions of ${\cal V}^{ns}$ and ${\cal A}^{ns}$ are also in full 
agreement. By this we mean that the coefficients of 
${\cal P}^{\cal I}_{(i) \mu }(p,q)$ for ${\cal O}^{ns}$~$=$~${\cal V}^{ns}$ and
${\cal A}^{ns}$ are equal for each $i$~$=$~$1$ to $6$ and thus establishes the 
equivalence with Larin's strategy in strictly four dimensions after 
renormalization.

One of the reasons for reviewing the non-singlet operator renormalization and
indicating an alternative way of defining the finite renormalization constant
is that it avoids the need to connect Green's functions for different
operators. For the singlet axial vector operator there is clearly no analogous
partner. Instead as highlighted in \cite{39,62} the associated finite
renormalization is derived from ensuring that the chiral anomaly is correctly
restored in four dimensional Green's function after renormalization. The 
non-singlet axial vector current is non-anomalous. Specifically the anomaly for
the singlet case is given by
\begin{equation}
\partial_\mu \left( \bar{\psi}^i \gamma^5 \gamma^\mu \psi^i \right) ~=~
a T_F \Nf \epsilon^{\mu\nu\sigma\rho} G^a_{\mu\nu} G^a_{\sigma\rho}
\label{chanom}
\end{equation}
where the right hand side can be written as a derivative of a gluonic current
and (\ref{vconns}) and (\ref{avconns}) become partners in this alternative view
of \cite{39,62}. In \cite{39} the two loop finite renormalization constant was 
computed and is given by
\begin{equation}
Z^{\mbox{\footnotesize{fin}}}_{{\cal A}^s} ~=~ 1 ~-~ 4 C_F a ~+~ \left[
22 C_F^2 - \frac{107}{9} C_F C_A + \frac{31}{9} C_F T_F \Nf \right] a^2 ~+~
O(a^3)
\label{zfinas}
\end{equation} 
which differs from (\ref{zfinans}) in the final two loop term. We have not 
reproduced this expression beyond one loop here due to a subtlety in 
(\ref{chanom}). This is to do with the fact that both operators of 
(\ref{chanom}) have first to be renormalized and the triangular mixing matrix 
of the operators determined, \cite{39,62}. Then to find 
$Z^{\mbox{\footnotesize{fin}}}_{{\cal A}^s}$ the anomaly equation itself, 
(\ref{chanom}), has to be inserted in a Green's function and evaluated in 
strictly four dimensions. In \cite{39} this was carried out by inserting in a 
gluon $2$-point function where the momentum flowing into one of the external 
gluon legs vanishes. With a total derivative present due to having to take the 
divergence of the current, the momentum flow into the operator cannot be 
nullified. This momentum configuration allows one to use the {\sc Mincer} 
algorithm, \cite{69}, which evaluates three loop massless $2$-point functions 
in dimensional regularization. A three loop calculation is in fact required to 
determine (\ref{zfinas}) due to the presence of the coupling constant $a$ on 
the right hand side of (\ref{chanom}). This means that the loop order of the 
anomaly Green's function and its coupling constant expansion are mismatched by 
one. Since this will also be the case for the symmetric point setup, the 
absence of the three loop master integrals for such Green's function 
evaluations means that that calculation cannot be carried out at present. This
also means that for the same reasons one cannot compute the two loop 
corrections to the singlet axial vector current conversion or matching function
calculated at one loop in \cite{32}. This function allows one to translate
results between two different renormalization schemes and in the case of
\cite{32} the respective schemes were $\MSbar$ and the symmetric momentum 
subtraction (SMOM) scheme. As the conversion function is computed from the 
singlet axial vector current renormalization constants in the two schemes then 
the finite renormalization constant for restoring four dimensional symmetries 
through the Larin method will also be needed for this. In turn this requires 
the finite parts of the three loop Green's function in both schemes. The 
absence of the three loop symmetric masters means that this cannot be carried 
out here. Having demonstrated, however, that the non-singlet axial vector 
finite $\MSbar$ renormalization constant consistently emerges in the symmetric 
point configuration, instead we merely accept the result (\ref{zfinas}) and use
it for our $\MSbar$ computations. In other words we include the extra graphs of
Figure \ref{fig2l} where ${\cal O}^{{\cal A}^s}$ is inserted. The 
$\gamma$-algebra is removed and the Lorentz tensor integrals evaluated by 
projection and the naive axial vector operator renormalization constant 
determined which is
\begin{equation} 
Z_{{\cal A}^s} ~=~ 1 ~+~ \left[ \frac{22}{3} C_F^2 + \frac{10}{3} C_F T_F \Nf 
\right] \frac{a^2}{\epsilon} ~+~ O(a^3) ~.
\label{zas}
\end{equation} 
This is in full agreement with the expression given in \cite{39}. Reproducing
this is a check on our different projection strategy since the $O(a^2)$ term
arises purely from the graphs of Figure \ref{fig2l}. After this naive 
renormalization the tensors of the Green's function are mapped to their four 
dimensional counterparts and then the finite renormalization constant
(\ref{zfinas}) is included. This ensures that the properties of the chiral 
anomaly are properly taken into consideration in the expression we have
computed for the Green's function of interest.

\sect{Results.}

Having discussed the computational strategy in detail we now present our 
results for the singlet axial vector Green's function. In order to give an 
indication of the structure of the final expression in four dimensions after 
the implementation of the Larin method we note that in the Landau gauge for 
$\Nf$~$=$~$3$ we have 
\begin{eqnarray}
\left. \Gamma^{{\cal A}^s}_\mu \right|_{\alpha=0}^{\Nf=3} &=& 
\left[ -~ 1
+ \left[
 \frac{8}{3}
+ \frac{16}{81} \pi^2
- \frac{8}{27} \psi^\prime(\third)
\right] a 
\right. \nonumber \\
&& ~+ \left. 
\left[
\frac{887}{9}
- \frac{12520}{27} s_3( \pisix )
+ \frac{10016}{27} s_3( \pitwo )
+ \frac{2504}{9} s_2( \pisix )
- \frac{5008}{9} s_2( \pitwo )
- \frac{1528}{27} \zeta_3
\right. \right. \nonumber \\
&& \left. \left. ~~~~~~
- \frac{3040}{243} \pi^2
- \frac{760}{729} \pi^4
+ \frac{1520}{81} \psi^\prime(\third)
+ \frac{32}{243} \psi^\prime(\third) \pi^2
- \frac{8}{81} \psi^\prime(\third)^2
\right. \right. \nonumber \\
&& \left. \left. ~~~~~~
+ \frac{91}{243} \psi^{\prime\prime\prime}(\third)
- \frac{9077}{4374\sqrt{3}} \pi^3
- \frac{626}{27\sqrt{3}} \ln(3) \pi
+ \frac{313}{162\sqrt{3}} \ln^2(3) \pi
\right] a^2 \right]
\gamma^5 \gamma_\mu
\nonumber \\
&& +~ \left[ \left[ 
\frac{32}{9}
+ \frac{32}{81} \pi^2
- \frac{16}{27} \psi^\prime(\third)
\right] a
\right. \nonumber \\
&& ~~~~~+ \left. 
\left[
\frac{2428}{27}
- \frac{6080}{9} s_3( \pisix )
+ \frac{4864}{9} s_3( \pitwo )
+ \frac{1216}{3} s_2( \pisix )
- \frac{2432}{3} s_2( \pitwo )
\right. \right. \nonumber \\
&& \left. \left. ~~~~~~~~~
- \frac{808}{9} \zeta_3
- \frac{5336}{243} \pi^2
- \frac{3680}{2187} \pi^4
+ \frac{2668}{81} \psi^\prime(\third)
+ \frac{128}{729} \psi^\prime(\third) \pi^2
\right. \right. \nonumber \\
&& \left. \left. ~~~~~~~~~
- \frac{32}{243} \psi^\prime(\third)^2
+ \frac{148}{243} \psi^{\prime\prime\prime}(\third)
- \frac{2204}{729\sqrt{3}} \pi^3
- \frac{304}{9\sqrt{3}} \ln(3) \pi
\right. \right. \nonumber \\
&& \left. \left. ~~~~~~~~~
+ \frac{76}{27\sqrt{3}} \ln^2(3) \pi
\right] a^2 \right] 
\left[ \gamma^5 \pslash p^\mu ~+~
\gamma^5 \qslash q_\mu \right] \frac{1}{\mu^2}
\nonumber \\
&& + \left[ 
\frac{16}{9} a
\right.
\nonumber \\
&& ~~~~+ \left. 
\left[ 
\frac{1214}{27}
- \frac{6800}{27} s_3( \pisix )
+ \frac{5440}{27} s_3( \pitwo )
+ \frac{1360}{9} s_2( \pisix )
- \frac{2720}{9} s_2( \pitwo )
\right. \right. \nonumber \\
&& \left. \left. ~~~~~~~~
- \frac{1280}{27} \zeta_3
- \frac{1400}{81} \pi^2
- \frac{880}{2187} \pi^4
+ \frac{700}{27} \psi^\prime(\third)
+ \frac{64}{729} \psi^\prime(\third) \pi^2
\right. \right. \nonumber \\
&& \left. \left. ~~~~~~~~
- \frac{16}{243} \psi^\prime(\third)^2
+ \frac{34}{243} \psi^{\prime\prime\prime}(\third)
- \frac{2465}{2187\sqrt{3}} \pi^3
- \frac{340}{27\sqrt{3}} \ln(3) \pi
\right. \right. \nonumber \\
&& \left. \left. ~~~~~~~~
+ \frac{85}{81\sqrt{3}} \ln^2(3) \pi
\right] a^2 \right] 
\left[ \gamma^5 \pslash q_\mu ~+~ 
\gamma^5 \qslash p_\mu \right] \frac{1}{\mu^2}
\nonumber \\
&& + \left[ \left[
\frac{32}{81} \pi^2
- \frac{16}{27} \psi^\prime(\third)
\right] a
\right. \nonumber \\
&& ~~~~~+ \left. 
\left[
80 s_3( \pisix )
- 64 s_3( \pitwo )
- 48 s_2( \pisix )
+ 96 s_2( \pitwo )
- \frac{112}{9} \zeta_3
+ \frac{64}{27} \pi^2
+ \frac{224}{729} \pi^4
\right. \right. \nonumber \\
&& \left. \left. ~~~~~~~~~
- \frac{32}{9} \psi^\prime(\third)
+ \frac{64}{243} \psi^\prime(\third) \pi^2
- \frac{16}{81} \psi^\prime(\third)^2
- \frac{4}{27} \psi^{\prime\prime\prime}(\third)
+ \frac{29}{81\sqrt{3}} \pi^3
\right. \right. \nonumber \\
&& \left. \left. ~~~~~~~~~
+ \frac{4}{\sqrt{3}} \ln(3) \pi
- \frac{1}{3\sqrt{3}} \ln^2(3) \pi
\right] a^2 \right] 
\gamma^5 \Gamma_{(3) \, p q \mu} \frac{1}{\mu^2} ~+~ O(a^3)
\end{eqnarray}
for $SU(3)$ where the restriction also means that the symmetric point 
conditions of (\ref{symmpt}) have been implemented and $\alpha$ is the 
covariant gauge parameter. As a further check on the computation we note that
the symmetry due to the interchange of the external legs of Figure \ref{figop} 
is present. This is the reason why pairs of basis tensors appear in the second 
and third terms and is consistent with the structure that emerged in the 
projection method used to evaluate $\Gamma^{{\cal A}^{ns}}$. The full 
expressions for the Green's function for arbitrary $\Nf$, linear covariant 
gauge parameter and general colour group are given in the attached data file. 
However to appreciate the relative size of the coefficients of each tensor for 
non-zero $\Nf$ and $\alpha$ we note that the Landau gauge numerical value of
$\left. \Gamma^{{\cal A}^s}_\mu \right|$ for $SU(3)$ is  
\begin{eqnarray}
\left. \Gamma^{{\cal A}^s}_\mu \right| &=& 
\left[ ~-~ 1
+ \left[ - 0.583194 \alpha + 1.624930 \right] a
\right. \nonumber \\
&& \left. ~
+ \left[ - 1.811999 \alpha^2 - 3.222901 \alpha + 3.889852 \Nf + 6.124832 
\right] a^2 \right]
\gamma^5 \gamma_\mu \nonumber \\
&&
+ \left[
\left[ 0.305695 \alpha + 1.472082 \right] a
\right. \nonumber \\
&& \left. ~~
+ \left[ 1.095408 \alpha^2 + 4.595782 \alpha - 7.001046 \Nf + 18.797491 
\right] a^2
\right] 
\left[ \gamma^5 \pslash p^\mu ~+~ \gamma^5 \qslash q^\mu \right] 
\frac{1}{\mu^2}
\nonumber \\
&&
+ \left[
\left[ 1.194584 \alpha + 1.777778 \right] a
\right. \nonumber \\
&& \left. ~~
+ \left[ 4.280593 \alpha^2 + 12.109050 \alpha - 7.550907 \Nf + 44.380585 
\right] a^2 
\right] 
\left[ \gamma^5 \pslash q^\mu ~+~ \gamma^5 \qslash p^\mu \right]
\frac{1}{\mu^2}
\nonumber \\
&&
+ \left[
-~ 2.083473 a
\right. \nonumber \\
&& \left. ~~
+ \left[ - 0.173623 \alpha^2 - 0.3484662 \alpha + 1.390610 \Nf - 39.787370 
\right] a^2
\right] 
\gamma^5 \Gamma_{(3) \, p q \mu} \frac{1}{\mu^2} \nonumber \\
&& +~ O(a^3) ~.
\end{eqnarray}
Clearly the coefficient of the $\gamma^5\gamma^\mu$ term of the two loop Landau
gauge Yang-Mills expression has the smallest magnitude in the $\MSbar$ scheme.
In order to gauge the effect the inclusion of the graphs of Figure \ref{fig2l}
have in comparison with the flavour non-singlet axial vector Green's function
it is instructive to compute the difference
$(\Gamma^{{\cal A}^{ns}}_\mu$~$-$~$\Gamma^{{\cal A}^s}_\mu)$. For example when 
$\Nf$~$=$~$3$ then for $SU(3)$ we have
\begin{eqnarray}
\left. \left[ \Gamma^{{\cal A}^{ns}}_\mu ~-~
\Gamma^{{\cal A}^s}_\mu \right] \right|^{\Nf=3} &=&
\left[ 
72
- \frac{1120}{3} s_3( \pisix )
+ \frac{896}{3} s_3( \pitwo )
+ 224 s_2( \pisix )
- 448 s_2( \pitwo )
- \frac{112}{3} \zeta_3
\right. \nonumber \\
&& \left. ~
- \frac{416}{27} \pi^2
- \frac{32}{81} \pi^4
+ \frac{208}{9} \psi^\prime(\third)
+ \frac{4}{27} \psi^{\prime\prime\prime}(\third)
- \frac{406}{243\sqrt{3}} \pi^3
\right. \nonumber \\
&& \left. ~
- \frac{56}{3\sqrt{3}} \ln(3) \pi
+ \frac{14}{9\sqrt{3}} \ln^2(3) \pi
\right] a^2 
\gamma^5 \gamma_\mu
\nonumber \\
&&
+ \left[
\frac{32}{3}
- \frac{1600}{3} s_3( \pisix )
+ \frac{1280}{3} s_3( \pitwo )
+ 320 s_2( \pisix )
- 640 s_2( \pitwo )
- \frac{160}{3} \zeta_3
\right. \nonumber \\
&& \left. ~~~
- \frac{800}{27} \pi^2
- \frac{64}{81} \pi^4
+ \frac{400}{9} \psi^\prime(\third)
+ \frac{8}{27} \psi^{\prime\prime\prime}(\third)
- \frac{580}{243\sqrt{3}} \pi^3
\right. \nonumber \\
&& \left. ~~~
- \frac{80}{3\sqrt{3}} \ln(3) \pi
+ \frac{20}{9\sqrt{3}} \ln^2(3) \pi
\right] 
\left[ \gamma^5 \pslash p^\mu ~+~ \gamma^5 \qslash q^\mu \right] 
\frac{a^2}{\mu^2}
\nonumber \\
&&
+ \left[ 
\frac{16}{3}
- \frac{640}{3} s_3( \pisix )
+ \frac{512}{3} s_3( \pitwo )
+ 128 s_2( \pisix )
- 256 s_2( \pitwo )
- \frac{64}{3} \zeta_3
\right. \nonumber \\
&& \left. ~~~
- \frac{416}{27} \pi^2
+ \frac{208}{9} \psi^\prime(\third)
- \frac{232}{243\sqrt{3}} \pi^3
- \frac{32}{3\sqrt{3}} \ln(3) \pi
\right. \nonumber \\
&& \left. ~~~
+ \frac{8}{9\sqrt{3}} \ln^2(3) \pi
\right] 
\left[ \gamma^5 \pslash q^\mu ~+~ \gamma^5 \qslash p^\mu \right]
\frac{a^2}{\mu^2}
\nonumber \\
&&
+ \left[
\frac{320}{3} s_3( \pisix )
- \frac{256}{3} s_3( \pitwo )
- 64 s_2( \pisix )
+ 128 s_2( \pitwo )
+ \frac{32}{3} \zeta_3
- \frac{64}{27} \pi^2
\right. \nonumber \\
&& \left. ~~~
+ \frac{64}{81} \pi^4
+ \frac{32}{9} \psi^\prime(\third)
- \frac{8}{27} \psi^{\prime\prime\prime}(\third)
+ \frac{116}{243\sqrt{3}} \pi^3
+ \frac{16}{3\sqrt{3}} \ln(3) \pi
\right. \nonumber \\
&& \left. ~~~
- \frac{4}{9\sqrt{3}} \ln^2(3) \pi
\right]
\gamma^5 \Gamma_{(3) \, p q \mu} \frac{a^2}{\mu^2} \,+\, O(a^3)
\label{singdiff}
\end{eqnarray}
which is independent of $\alpha$ or
\begin{eqnarray}
\left. \left[ \Gamma^{{\cal A}^{ns}}_\mu ~-~
\Gamma^{{\cal A}^s}_\mu \right] \right|^{\Nf=3} &=&
\left[ 10.960779 \mu^2 \gamma^5 \gamma_\mu
- 17.013282 \left[ \gamma^5 \pslash p^\mu ~+~ \gamma^5 \qslash q^\mu \right] 
\right. \nonumber \\
&& \left. ~
- 14.060128 \left[ \gamma^5 \pslash q^\mu ~+~ \gamma^5 \qslash p^\mu \right]
- 4.856554 \gamma^5 \Gamma_{(3) \, p q \mu} 
\right] \frac{a^2}{\mu^2} \nonumber \\
&& +~ O(a^3)
\end{eqnarray}
numerically.

Given this in order to appreciate the effect of the finite renormalization 
associated with the chiral anomaly in a clear way, it is instructive to plot 
(\ref{singdiff}) as a function of a dimensionless momentum variable. To achieve
this we recall that solving the two loop $\beta$-function as a function of the 
squared momentum $Q^2$ and the QCD $\Lambda$ parameter gives
\begin{equation}
a_2(Q,\Lambda) ~=~ \frac{1}{b_0 L} \left[ 1 - \frac{b_1 \ln (L)}{b_0^2 L}
\right]
\end{equation}
where the $\beta$-function coefficients are, \cite{70,71,72,73},
\begin{eqnarray}
b_0 &=& \frac{11}{3} C_A ~-~ \frac{4}{3} T_F \Nf \nonumber \\
b_1 &=& \frac{34}{3} C_A^2 ~-~ 4 T_F C_F \Nf ~-~ \frac{20}{3} T_F \Nf C_A
\end{eqnarray}
in the $\MSbar$ scheme and we use the shorthand 
\begin{equation}
L ~=~ \ln \left( \frac{Q^2}{\Lambda^2} \right) 
\end{equation}
for the logarithm which is present. In Figure \ref{figcomp} we have plotted the
two loop coefficient of $\gamma^5\gamma^\mu$ for $\Gamma^{{\cal A}^{ns}}$ and 
$\Gamma^{{\cal A}^s}$ for several values of $\Nf$ where the dimensionless
variable $x$ is defined by $x$~$=$~$Q^2/\Lambda^2$. At $x$~$=$~$3$ for
example, the difference between the coefficients of this particular tensor
range from around $2\%$ to $7\%$ at $x$~$=$~$3$ from $\Nf$~$=$~$3$ to $6$
respectively. As an alternative to the explicit values Figure \ref{figdiff} 
shows the difference for the same values of $\Nf$ against $x$. Clearly at very 
high energy the discrepancy tends to zero.  

{\begin{figure}[ht]
\begin{center}
\includegraphics[width=7.7cm,height=7.7cm]{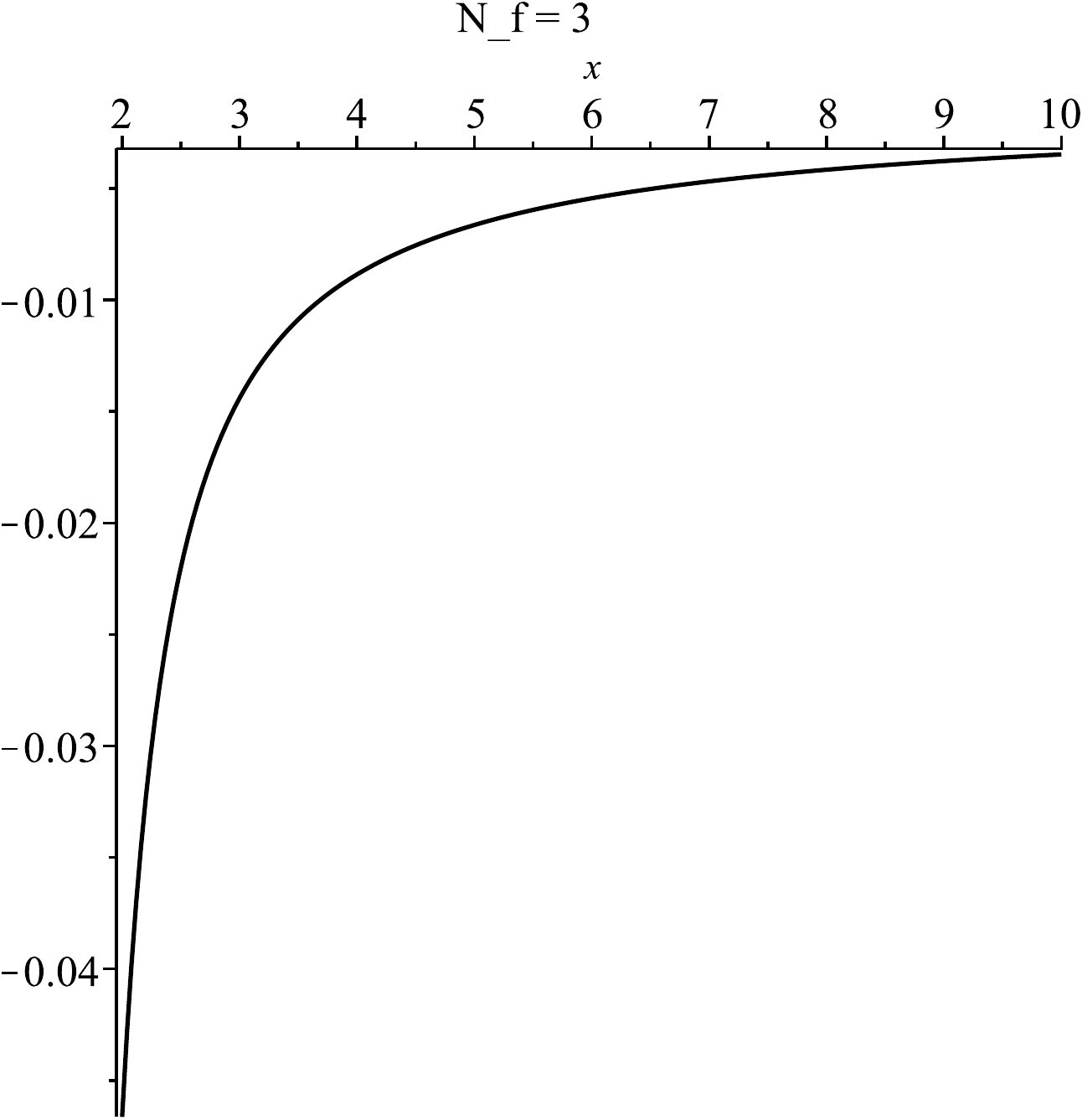}
\quad
\includegraphics[width=7.7cm,height=7.7cm]{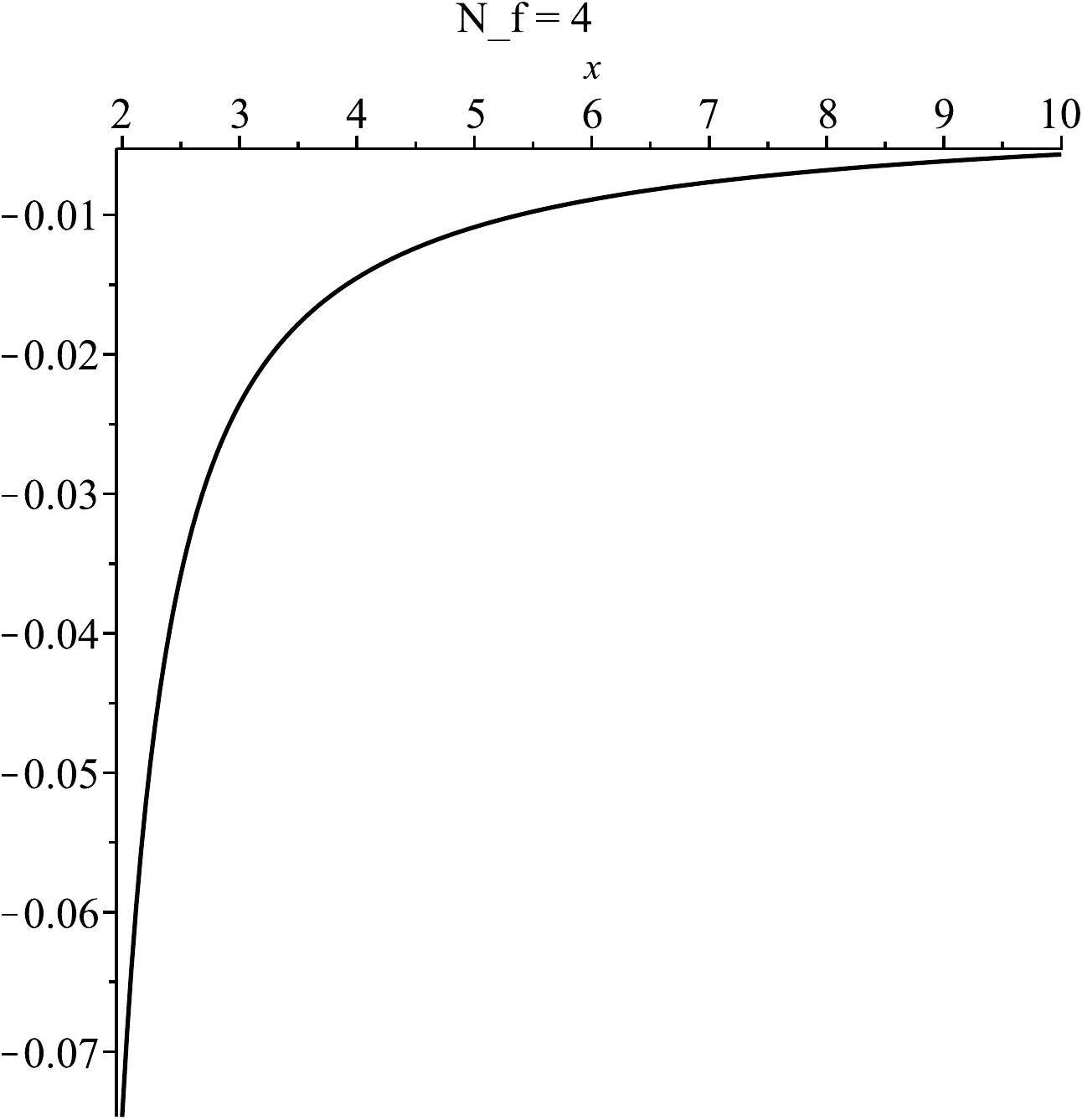}

\vspace{0.8cm}
\includegraphics[width=7.7cm,height=7.7cm]{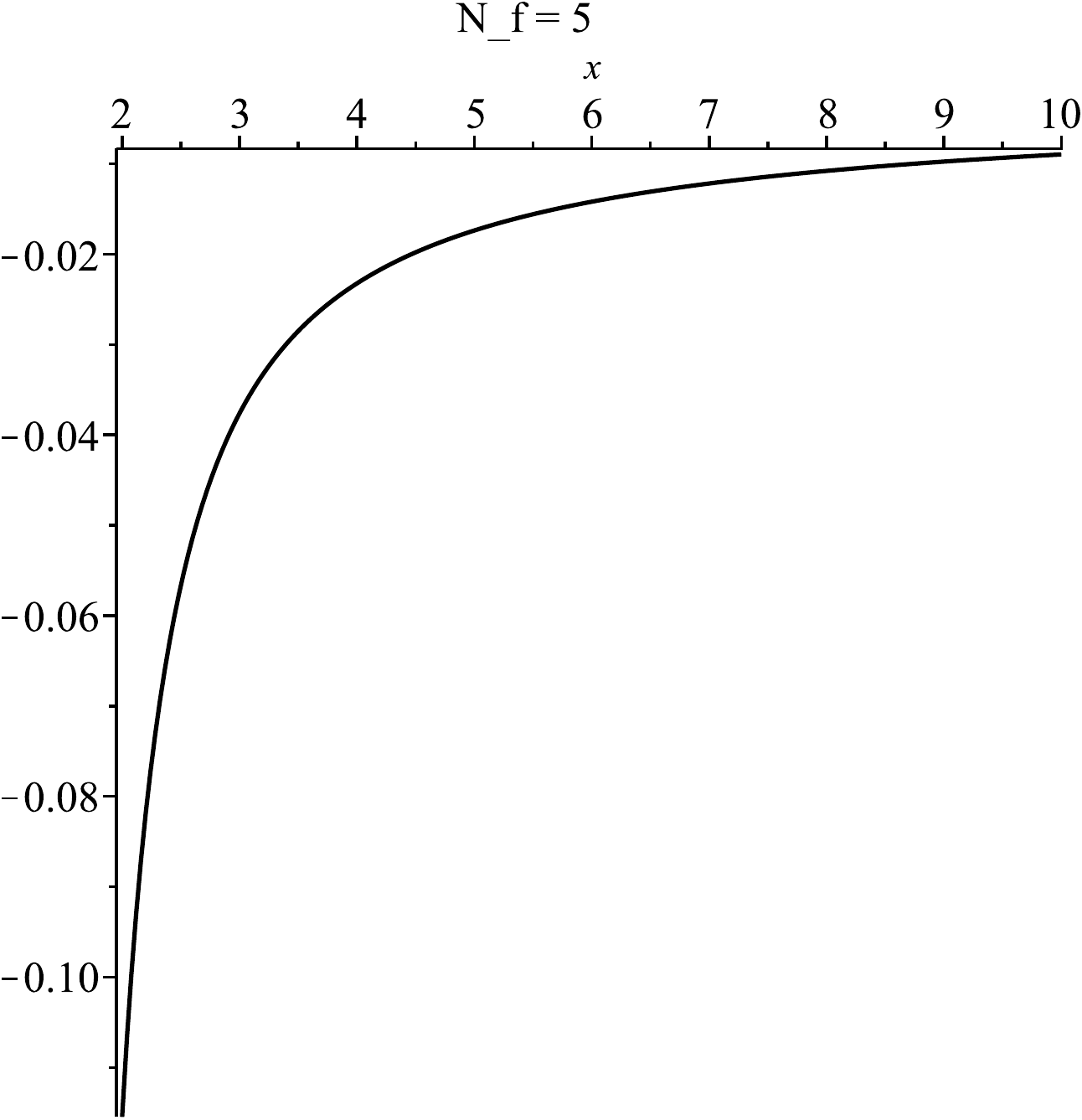}
\quad
\includegraphics[width=7.7cm,height=7.7cm]{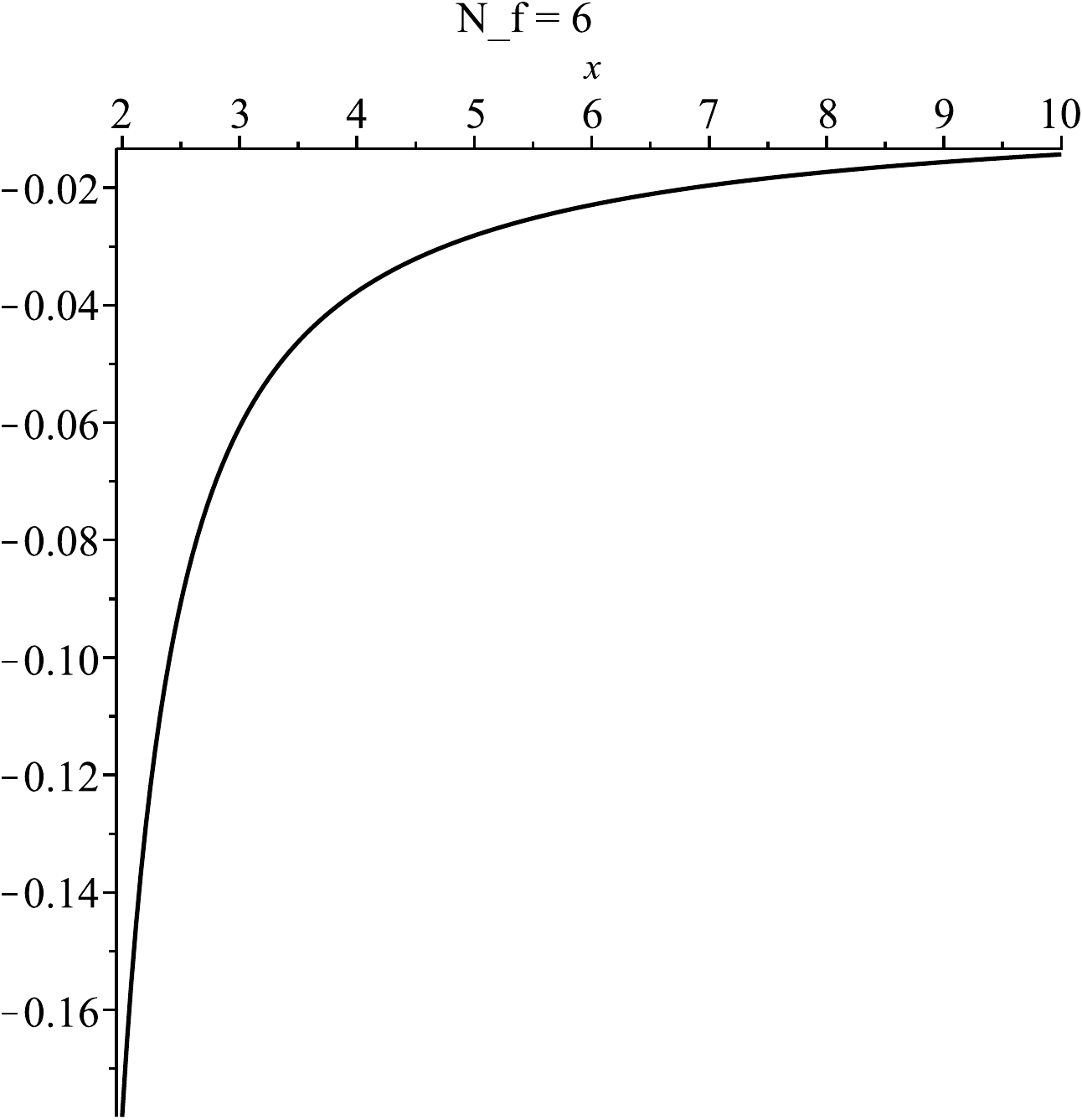}

\end{center}
\caption{Two loop coefficient of $\gamma^5\gamma^\mu$ in
$\left[ \Gamma^{{\cal A}^{ns}}\right.$~$-$~$\left.\Gamma^{{\cal A}^s}\right]$ 
for $\Nf$~$=$~$3$ to $6$.}
\label{figdiff}
\end{figure}}

\sect{Discussion.}

We have evaluated the two loop quark $2$-point function with the flavour 
singlet axial vector current inserted at non-zero momentum at the 
non-exceptional symmetric point. The main feature of this result is that unlike
the one loop Green's function we have had to ensure that our construction is 
consistent with the non-conservation of the singlet axial current due to the 
chiral anomaly. This was a non-trivial task since the use of dimensional
regularization means that the purely four dimensional $\gamma^5$-matrix has to
reconciled. To achieve this we implemented the Larin method, \cite{39,62}, 
which was originally developed for the quark bilinear operators and founded 
upon \cite{40,41,42,62}. However, the finite renormalization constant 
associated with each $\gamma^5$ dependent operator was determined in 
\cite{39,62} for an exceptional momentum configuration. Here we have confirmed 
that these expressions for the flavour non-singlet operators are independent of
the subtraction point for the renormalization scheme that we have used which is 
$\MSbar$. If one were to use another scheme such as the momentum subtraction 
scheme of \cite{43,44} then a different finite renormalization constant would 
emerge. For the flavour singlet axial vector case one would have to extend our 
two loop computation to three loops at the symmetric point to verify Larin's 
two loop finite renormalization constant for that operator. This is due to the 
non-conserving part of the anomaly equation being proportional to the coupling 
constant. 

Concerning the comparison of the axial vector flavour non-singlet and singlet 
corrections at two loops, the difference at a representative momentum scale is 
of the order of a few percent depending on the value of $\Nf$. It is not clear 
whether such a value would make a significant difference to the analysis 
already carried out in \cite{27,32} for instance. It would depend on whether 
there is a clear signal in the lattice measurements to differentiate between 
the various Green's functions. Of course this situation could be improved by 
extending to three loops which is the next natural step in the process. Aside 
from the absence of the three loop master integrals at the symmetric point, one
would still have to be careful with the treatment of $\gamma^5$ in dimensional 
regularization. For instance, the finite renormalization constant for the axial
vector current is known to three loops for the non-singlet case but only two 
loops for the singlet case. To put the latter on the same level as the former 
would require a four loop evaluation. While the computational tools are 
available through the development of the {\sc Forcer} package, \cite{74,75}, 
the treatment of $\gamma^5$ at four loops possibly requires detailed care. 
Recently there has been progress in this direction through the careful 
determination of the four loop Standard Model $\beta$-functions, \cite{76}. 
This was achieved by ensuring general quantum field theory consistency 
conditions were satisfied for the $\gamma^5$ sector. By contrast for any 
treatment of $\gamma^5$ that uses the Larin approach where spinor traces 
involve a large number of $\gamma$-matrices it has been noted in \cite{77} that
one has always to be fully aware of potential hidden evanescence issues. In our
case since we use (\ref{gamn}) in dimensional regularization the latter point 
would need to be accommodated at three and higher loops.

\vspace{1cm}
\noindent
{\bf Acknowledgements.} This work was supported by a DFG Mercator Fellowship.
The author thanks both Dr J. Green and Dr P.E.L. Rakow for discussions and 
comments on a draft, and to the former for indicating the necessity of this 
calculation. The Mathematical Physics Group at Humboldt University, Berlin is
thanked for its hospitality where this work was initiated and partly carried 
out. Figures $1$ and $2$ were prepared with the {\sc Axodraw} package, 
\cite{78}.

\end{document}